\documentclass[aps,prb,twocolumn,showpacs]{revtex4-1}
\usepackage{amsmath,amssymb,graphicx,epsfig,color,bbm}

\def\nn{\nonumber}

\begin{document}

\title{Precision of electromagnetic control of a quantum system}

\author{Ching-Kit Chan and L. J. Sham}
\affiliation{Center for Advanced Nanoscience, Department of Physics, University of California San Diego, La Jolla, California 92093-0319, USA}
\date{\today}

\begin{abstract}

Coherent control of a quantum system is limited both by the decoherence due to environment and the quantum nature of the control agent. The high fidelity of control demanded by fault tolerant quantum computation and the intrinsic interest in nonclassical effects from the interplay between control and dissipation are motivations for a detailed study of the interaction dynamics between the quantum system and the macroscopic environment and control agent. We present a detailed time evolution study of a two-level system interacting with a laser pulse and the electromagnetic vacuum in the multimode Jaynes-Cummings model. A diagrammatic formalism allows easy identification of coherent dynamics and relaxation of the two-level system. We demonstrate a computational method of dynamics with precise error bounds for fast operations versus slow decoherence, spanning the Markovian and non-Markovian regimes. Comparison against an exact model solution of our results with existing approximations of the master equation shows the lack of accuracy in the latter.

\end{abstract}


\maketitle

\section{Introduction}\label{sec_introduction}

An open system, i.e., a small quantum object in the presence of a macroscopic environment, presents a fundamental problem in quantum mechanics and its applications. We wish to address here the problem of the environment with the dual function of decoherence and control of the quantum object. For practical purposes, both experiments on coherent processes and quantum technology require a small parameter $t_0/T_2$ in the time scale of the control duration $t_0$ being much smaller than the decoherence time $T_2$. A paradigmatic system for this problem is the interaction of a two-level system (TLS) with the quantized electromagnetic fields.\cite{jaynes63} In terms of the TLS-photon interaction strength $g$ in units of frequency, the controlled TLS process (the Rabi rotation) is a strong coupling process with $g|\alpha| t_0\sim O(1)$ using a coherent photon state $|\alpha\rangle$ with a large mean number $|\alpha|^2$ of photons while the long decoherence time is a weak coupling process $t_0/T_2 \ll 1$ within the control time. In particular, the high fidelity of the operation to an error threshold between $10^{-3}$ and $10^{-4}$ in fidelity demanded by fault tolerant quantum computation\cite{aliferis09,gottesman09} sets the bar for high accuracy in the theory of the open system. While the TLS open system problem has been much studied, we posit that the specific additive problem of the control and decoherence processes remains.

The decoherence problem of a TLS in a spin bath and its suppression under a classical control have been extensively studied,\cite{khodjasteh05,grace07} but the noise due to a quantized optical control was not taken into account. Barnes and Warren demonstrated the decoherence induced by the back action from the TLS to the electromagnetic control.\cite{barnes99} For a Markovian system, this problem can be solved using the optical Bloch equation\cite{berman11}. However, in many non-Markovian systems with a structured environment, say nanocavities\cite{yoshie04} and photonic band gap materials\cite{lodahl04}, the optical Bloch analysis no longer applies. Because of the large Hilbert space of the multimode TLS-field Hamiltonian, an exact diagonalization of the problem is impractical.\cite{swain72} Therefore, a quantum theory suitable for a non-Markovian open system with high accuracy set for our problem is needed.

In this paper, we develop a formalism to solve the multimode Jaynes-Cummings (JC) model under a coherent light pulse with arbitrary pulse shape in the limit $t_0\ll T_2$, relevant to the problem posed. The exact non-unitary evolution of the TLS is expressed in terms a time evolution forward and a reverse evolution backward of emitted and absorbed photons conditioned on the TLS down and up transitions. The photon dynamics are evaluated by the field theoretic perturbation series. The diagrammatic structure allows an explicit identification of the photons' role in coherent dynamics and dissipation process of the TLS. All perturbation terms in a coherent dynamics segment are summed and the dissipative processes are expanded in powers of a small parameter $(t_0/T_2)^{\gamma}$ where $\gamma$ depends on the photon correlation time $\tau_c$ and ranges between 1 (the Markovian limit) and 2 (the non-Markovian limit). The process gives a practical numerical computation procedure for a given error limit in powers of the small parameter. We illustrate the computation for a short-time quantum operation of the TLS by including all the relevant quantum processes within the first order error bound. We identify the precise origins of the interference effects between the control and the dissipation in the evolution processes.

The master equation (ME) approach of treatment of the open quantum system has been enormously important for quantum optics \cite{carmichael} and for quantum information.\cite{wiseman} We have made a comparison of our theory with several prominent approximations extant in the ME approach\cite{breuer02} for a model problem involving the interference effect between control and decay which has an exact solution and a semiclassical one. The comparison results show that in the non-Markovian regime, the driving forces for the coherent control and the dissipation are not additive and that the ME approximations are closer to the semiclassical results than the quantum results, not meeting the stringent accuracy requirement. Perhaps, the comparison results would stimulate effort to refine the ME approximations.

The outline of the rest of the paper is as follows. In Section~\ref{sec_formalism}, we develop a field-theoretic solution for the JC model in the presence of the multimode photonic field, elucidating the coherent and dissipative components from the diagrammatic structure. In Sec.~\ref{sec_non-Markovian dynamics}, a detailed analysis of the effect of vacuum decoherence on the control precision of the TLS is given in the non-Markovian regime. Sec~\ref{sec_comparison} compares the diagrammatic solution with the ME approximations. Sec.~\ref{sec_conclusion} summarizes and the appendices add some technical details.


\section{TLS-photons interaction dynamics}\label{sec_formalism}

By the field theoretic techniques, we express the evolution operator of the whole system (the TLS and the photons) as an infinite perturbation series in terms of the spin and photon operators. Then, we evaluate the matrix elements of the spin exactly, resulting in a diagrammatic series of the photon operators only. By the Wick's theorem on the photon operators, we build a perturbative solution to the non-equilibrium problem of the dynamics of the laser photons and the TLS in the bath of the electromagnetic vacuum. We find a controlling small parameter, $(t_0/T_1)^{\gamma}$, defined in the Introduction and detailed below, for the perturbation series. The approach of removing the spin operators first stands in contrast to the standard ME approach\cite{breuer02} which traces out the photonic environment first and then solves the equation of motion of the TLS, and which lacks error bounds for most of its approximations.

\subsection{TLS transformation by coherent photon state}

We start with the canonical multimode JC Hamiltonian:\cite{jaynes63, shore93}
\begin{eqnarray}
H &=& H_0 + V, \label{eq-h} \\
\text{where} \quad
H_0 &=& \frac{1}{2}\omega_0\sigma_z + \sum_k \omega_k a_k^{\dagger} a_k, \nn \\
V   &=& \sum_k g_k \Big (\sigma_{+}a_{k}+\sigma_{-}a_{k}^{\dagger}    \Big ).
\end{eqnarray}
$H_0$ contains the bare Hamiltonian of the TLS of energy splitting $\omega_0$ with the Pauli operator $\sigma_z$ and the photons of energy $\omega_k$ with creation operator $a_k^{\dagger}$. $V$ is the TLS-photon interaction with the coupling constant $g_k$, presented in the rotating wave approximation (RWA) which is justified in Appendix~\ref{sec_appendix_A}. While the single-mode JC may be used to treat the laser pulse by making the coupling time-dependent $g(t)$, the multimode extension \cite{berman11} is more suited for our purpose of investigating the joint quantum effects of the light control of TLS and the dissipation.

The composite system of the TLS and photons is given by a product initial wavefunction, $|\Psi(0)\rangle =\left[\sum_s c_{s} |s\rangle \right]|\boldsymbol{\alpha}\rangle $, where $s=\pm$ denotes the two states of the TLS and $\boldsymbol{\alpha}=( \alpha_{k_1},\alpha_{k_2},\ldots)$ denotes the multimode coherent state. The reduced density matrix $P_{s_f,s_f'}(t)$ is expressed in terms of the transformation matrix conditioned on the initial photon state:
\vspace{-0.066in}
\begin{eqnarray}
P_{s_f,s_f'}(t) &=&  \sum_{s,s'} c_{s}c_{s'}^* p_{s_f,s_f';s,s'}(t,\boldsymbol{\alpha}) \\
p_{s_f,s_f';s,s'}(t,\boldsymbol{\alpha}) &=& \langle\boldsymbol{\alpha} | \langle s'| U^{\dagger}(t) |s_f' \rangle \langle s_f| U(t) |s\rangle |\boldsymbol{\alpha} \rangle \nn \\
&& \times \ e^{i(s_f'1-s_f1)\omega_0 t/2}.
\label{eq_transformation_matrix}
\end{eqnarray}
In the interaction picture, the time evolution operator and the interaction are:
\begin{eqnarray}
U(t) &=& T \exp \Big[ -i \int_0^t dt' V(t') \Big], \nn \\
V(t_l) &=& \sigma_+ A_l + \sigma_- A_l^\dagger, \\
\label{eq-v}
\text{where~~}
A_{l}&=&\sum_{k}g_{k}a_{k}e^{i\Delta_k t_{l}},\nn \\
A_{l}^{\dagger}&=&\sum_{k}g_{k}a_{k}^{\dagger}e^{-i\Delta_k t_{l}},
\label{eq-u4}
\end{eqnarray}
and $\Delta_k =\omega_{0}-\omega_{k}$
is the detuning of the $k$ mode. This form of transformation of the reduced density matrix visibly retains the quantum nature of the evolution of the composite system and is easily reduced to the problem of the expectation value of the electromagnetic field operators for the initial photon state.
The initial product state may be generalized to any composite state $\left.|\Psi(0)\right\rangle =\sum_s \int \mathbf{D}\boldsymbol{\alpha}c_{s ,\boldsymbol{\alpha}} \left.|s\right\rangle \left.|\boldsymbol{\alpha}\right\rangle$, where $\int \mathbf{D}\boldsymbol{\alpha}=\int \frac{d^2\alpha_{k_1}}{\pi}\frac{d^2\alpha_{k_2}}{\pi}\ldots $

In the perturbation series of the evolution operator, the TLS state is flipped up or down by a series of interaction, leaving only the corresponding photon operators,
\begin{align}
\langle \pm | U(t) | \pm \rangle &= \sum_{n=0}^\infty (-i)^{2n} \int_0^t D^{2n}t~X_{2n}^{\pm} ,\label{eq-u} \\
\langle \mp | U(t) | \pm \rangle &= \sum_{n=0}^\infty (-i)^{2n+1} \int_0^t D^{2n+1}t~X_{2n+1}^{\pm},
\label{eq-u6} \\
\text{where} \quad
\int_0^t D^n t &= \int_0^t dt_n ... \int_0^{t_3} dt_2 \int_0^{t_2} dt_1, \nn \\
X^+_{2n} &= A_{2n} A_{2n-1}^\dagger~...~A_2 A_1^\dagger ,\nn \\
X^+_{2n+1} &= A_{2n+1}^\dagger X_{2n}^+ ,\nn \\
X^-_{2n} &= A_{2n}^\dagger A_{2n-1}~...~A_2^\dagger A_1 ,\nn \\
X^-_{2n+1} &= A_{2n+1} X_{2n}^-.
\label{eq_X}
\end{align}

\subsection{TLS evolution in terms of photon processes}

The task is reduced to evaluate the transformation matrix in Eq.~(\ref{eq_transformation_matrix}) in terms of the series expansion of the unitary operator and its inverse in Eq.~(\ref{eq-u}) or (\ref{eq-u6}) that consist of photonic components in Eqs.~(\ref{eq_X}). The formulation is exact so far. The evaluation of the series is simplified by the Wick's theorem \cite{wick50} (see Appendix~\ref{sec_app_BNEW}). Each series term in Eq.~(\ref{eq-u}) is of the form $X^{\pm\dagger}_mX^\pm_{m'}$ from Eqs.~(\ref{eq_X}), a product of several photon operators $A_i$ and$ A_{j}^\dagger$, which the Wick's theorem resolves into a sum of terms composed of a normal product and a number of pairs of contractions. The matrix element of each normal product between two coherent states is simply products of scalars from the substitution of $a_k\rightarrow \alpha_k$ acting on the coherent ket vector and $a_k^{\dagger}\rightarrow \alpha_k^*$ acting on the bra vector.
The contractions from Eq.~(\ref{eq-contraction}) are
\begin{eqnarray}
\langle A_i ^\dagger A_{j}\rangle &=& 0,  \label{eq-a0}\\
\langle A_i A_{j}^\dagger \rangle &=&  K(t_i-t_j)= \sum_k |g_k|^2 e^{i\Delta_k (t_i - t_j)}. \label{eq-aa}
\end{eqnarray}

\begin{figure}[t]
\begin{center}
\includegraphics[angle=0, width=0.9\columnwidth]{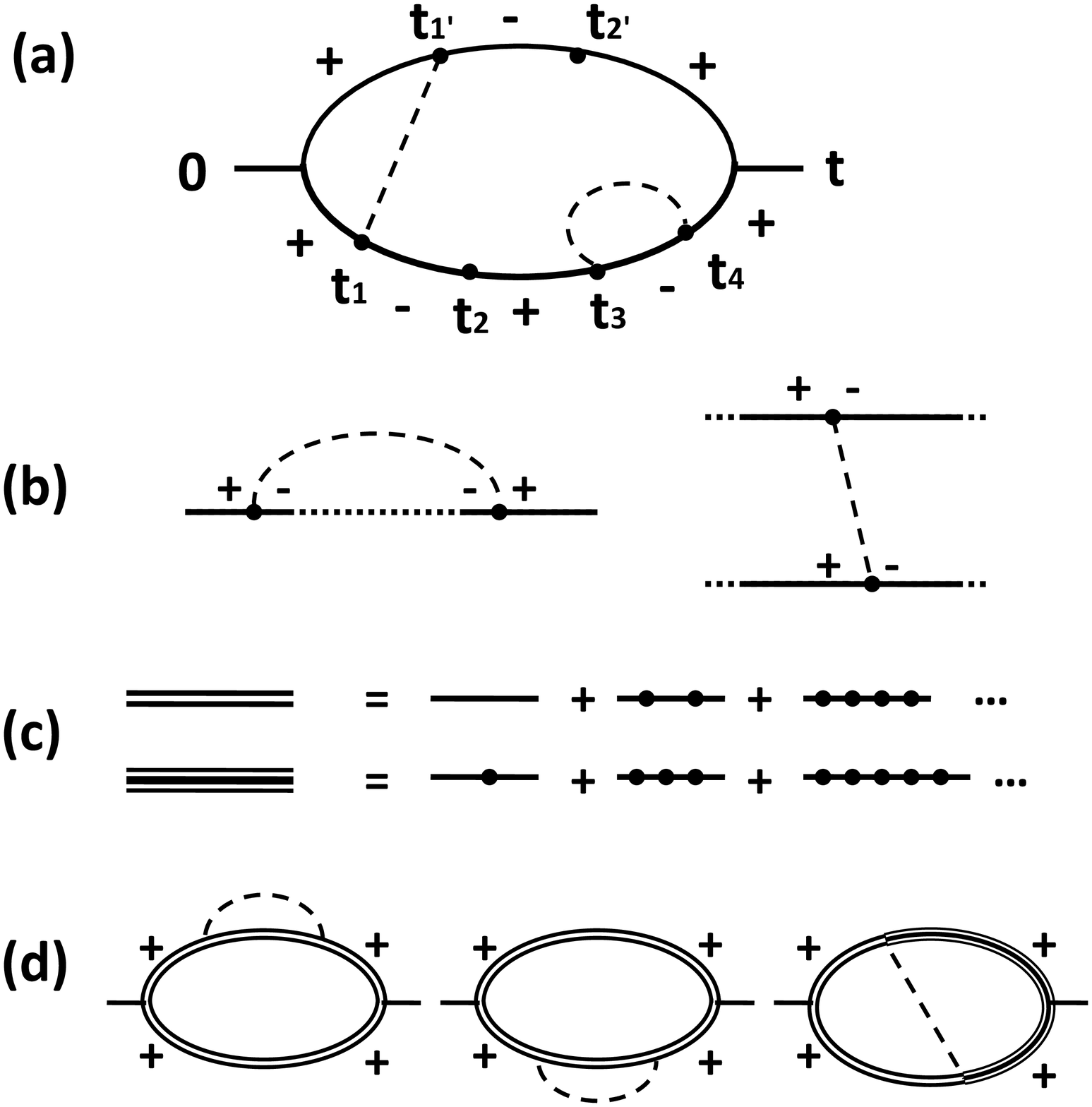}
\caption{Diagrammatic representations of perturbation terms of the transformation matrix $p_{s_f,s_f';s,s'}(t,\{\alpha\}) $. (a) An example of six photon operators with two contractions. The two short lines labeled $0,t$ are the time limits of the integrals. Each dot labeled with a time variable $t_n$ represents one photon operator, the lower solid arc being time ordered and the upper arc anti-time-ordered. A dashed line stands for one contraction between two photonic operators. The uncontracted dots form a normal product of the photon operators for the matrix element of the coherent states. The $\pm$ sign denotes the state of the TLS at different times. (b) The only possible contractions drawn between photons on the same or opposite time lines. (c) The dressed line by a sum of all even or all odd numbers of photons interacting with TLS without contraction. (d) Three diagrams that contains only a single contraction for the transformation matrix $p_{++;++}$. They are the leading contributions to the control noise.}
\label{fig_1}
\end{center}
\end{figure}

Fig.~\ref{fig_1} illustrates the diagrammatic representation of each term in the series expansion of the transformation matrix and some partial summations of subseries. Fig.~\ref{fig_1}(a) shows a typical term. The rules are: (i) the initial and final TLS states represented at times $0$ and $t$, (ii) the interactions with the TLS denoted by dots in the counterclockwise loop for the time-ordered photon operators from time $0\to t$ and then in anti-time order back $t \to 0$, (iii) the appropriate TLS states between dots labeled as $\pm$ and (iv) all possible contractions (dashed lines) either on the same time line or between the opposite time lines. Fig.~\ref{fig_1}(b) shows the only two possible types of contractions because of Eqs.~(\ref{eq-a0},\ref{eq-aa}). If the two ends of a solid line segment have the same (opposite) TLS states, the segment is dressed by an even (odd) number of photons whose series sum is depicted by a double (triple) line, see Fig.~\ref{fig_1}(c). Fig.~\ref{fig_1}(d) provides an example of the transformation matrix $p_{++,++}$ with a single contraction and dressed states. We stress that these diagrams are first order in contraction, but infinite order in the coherent interaction.

\subsection{The key results}

The quantum treatment of the time evolution of the TLS in terms of a series of photon contractions yields a number of notable results. The series expansion in the TLS-photon interaction permits identification of the physical processes. Well-known are the all-dots terms uninterrupted by dotted lines, as in the first term of Eq.~(\ref{eq_wick}) whose sum yields the Rabi rotation and the complete contraction pairs in the last term of the same equation whose sum yields the relaxation due to the electromagnetic vacuum (see Appendix~\ref{sec_appendix_C}). In the mixed terms, the dots between the ends of contractions may still be summed as coherent dynamics of the TLS. The summation is necessary because along the dressed line, each photon gives a term of the order $g|\alpha| t_0$ for a pulse of duration $t_0$. These series expressions can be summed exactly and are given in Appendix~\ref{sec_appendix_C}. The coexistence of the coherent dynamics and relaxation yields an effect which is extra to the sum of the two processes, as will be shown next. This is also clear in the nature of the contraction between the evolution and its inverse shown in Fig.~\ref{fig_1}(a) and (d). Note how, without the manual separation of the photon Hamiltonian into a control part and a bath part, the method produces the dissipation and the Rabi rotation. More importantly, it contains quite explicitly the interference effects between the two processes.

A useful result for our stated purpose of studying the fidelity of the control process to high accuracy is the finding of a small parameter for the expansion. After the construction of the dressed lines of the coherent processes, the expansion of the contraction functions is a perturbation series in powers of $(t_0/T_1)^\gamma$ in terms of the operation time $t_0$ and the decay time $T_1$ (see Appendix~\ref{sec_appendix_C} for the decoherence time $T_2=2^{1/\gamma}T_1$). The decay in each contraction line is $\sim \int \int dt_i dt_j K(t_i-t_j)\sim O[(t_0/T_1)^\gamma] $, where the parameter $\gamma$ between $(1, 2)$ depends on the shape of the photon DOS. Therefore, this method provides an excellent evaluation of the control noise in the regime $t_0 \ll T_1$, while the area of the pulse $\sim g|\alpha| t_0 \sim O(1)$. The one-contraction terms in Fig.~\ref{fig_1}(d) together give the consistent result to the first order contribution of the control noise which will be evaluated next. The consistency of the three diagrams comes from the differentiation of a self-energy diagram with one contraction resulting in three terms, as in the Ward identity in field theory. \cite{ward50}

\section{Relaxation and control fidelity}\label{sec_non-Markovian dynamics}

The quantum effects between the coherent control and the vacuum decoherence depend crucially on the contraction, which, Eq.~(\ref{eq-aa}) shows, depends on the materials properties of the TLS and the photon confinement in the form of the DOS of the photonic field weighed by the interaction mode dependence. To show explicitly the dependence of the control dynamics on the weighted DOS, it is convenient to model it as a Gaussian:
\begin{eqnarray}
\rho(\omega)=\sum_{k}g_{k}^{2}\delta(\omega-\omega_{k})=\frac{g^{2}\tau_{c}}{2\sqrt{\pi}}e^{-\tau_{c}^{2}(\omega-\omega_{0})^{2}/4},
\label{eq_rho_omega}
\end{eqnarray}
where g is the average coupling strength and $\tau_{c}$ describes the correlation time of multimode light in the presence of a TLS. The qualitative results, such as the concept of the exponent $\gamma$ in decay, are unchanged for a general DOS that includes a correct behavior for $\omega \rightarrow 0^+$. When $\tau_{c}\rightarrow0$, the broadband DOS yields $T_1=1/ \sqrt{\pi}g^2\tau_c$; while as $\tau_{c}\rightarrow\infty$, the single mode scenario pertains. Under this Gaussian DOS, Eq.~(\ref{eq_contraction}) shows that the contraction function is also a Gaussian, i.e., $K(t) = g^2 e^{-t^2/\tau_c^2}$.

For a broadband DOS, e.g., in free space, the system is Markovian and the result is equivalent to that from solving the optical Bloch equations.\cite{berman11} For an extremely narrow DOS, single-mode cavity quantum electrodynamics dominates. In this section, we use a Gaussian DOS whose variable width causes the decay of the upper state of the TLS to have a dependence of $\ln P_{++}(t) \propto -(t/T_{1})^{\gamma}$ characterized by the exponent $\gamma$ and use the change of the system from the exponential decay ($\gamma=1$) in the broad DOS limit to a Gaussian decay ($\gamma=2$) in the narrow DOS region to show the emergence of the quantum effects of interference between the laser control and the vacuum decoherence. The interference actually has a beneficial effect on the fidelity of the quantum operation on the TLS.

An example of a relevant physical system is a multimode finite Q cavity system with bandwidth $\sim 1/t_0$. For a photonic crystal nanocavity\cite{yoshie04} with $Q\sim 10^{4}$--$10^{5}$ and $\omega_0 \sim 10^{15}~\text{Hz}$, the estimated value of $\tau_c$ is $\sim 100~\text{ps}$ and non-Markovian features could be observable for $t_0\sim\tau_c$. Then, the small parameter condition for a fidelity of $0.9999$ could be relaxed from $t_0/T_1\sim 10^{-4}$ to $10^{-2}$.

\subsection{Non-markovian relaxation for a Gaussian DOS}\label{subsec_nonMarkov_relaxation}

\begin{figure}[t]
\begin{center}
\includegraphics[angle=0, width=0.9\columnwidth]{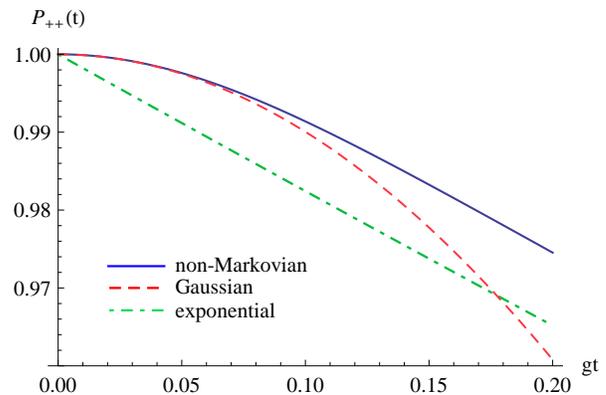}
\caption{(color online). Relaxation of an initially excited TLS state in vacuum using a Gaussian photon DOS with $g\tau_c=0.1$. The decoherence of the TLS (solid curve) is different from the purely Gaussian (dashed) or Markovian (dash-dotted) decay.}\label{fig_2}
\end{center}
\end{figure}

The short time decoherence behavior of the non-Markovian multimode JC system in vacuum is given by diagrams with only one contraction (see Fig.~\ref{fig_8}(b)). For an initially excited TLS, we have
\begin{align}
\ln[P_{++}(t)] \approx & -2\int_{0}^{t}dt_{1}\int_{0}^{t_{1}}dt_{2} K(t_1-t_2).
\end{align}
An illustration using $g\tau_c=0.1$ is presented in Fig.~\ref{fig_2}. The result reveals that for a finite value of $\tau_{c}$, the system evolves from a non-Markovian Gaussian dependence (well-known in the onset of the collapse and revival phenomena,\cite{shore93, rempe87, meekhof96, brune96}), $P_{++}(t)\approx e^{-(gt)^{2}}$, to the Markovian exponential decay, $P_{++}(t)\approx e^{-t/T_{1}}$. The crossover takes place at $t\sim\tau_{c}$. For short time, the contraction function is almost flat and the decay resembles the single mode case. As time increases and exceeds $\tau_c$, the Gaussian contraction function approaches the broadband Markovian limit. Note that both the Markovian and single mode approximations overestimate the decoherence of the TLS.

\begin{figure}[th]
\begin{center}
\includegraphics[angle=0, width=0.9\columnwidth]{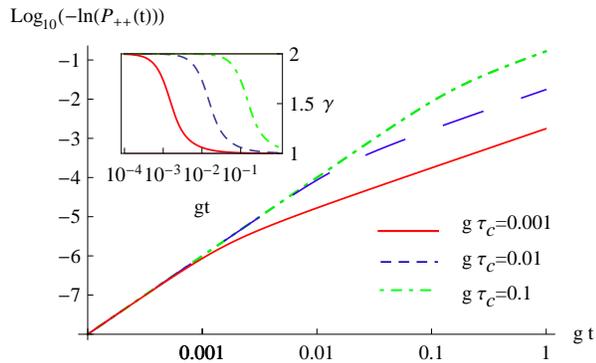}
\caption{(color online). Plots of $\log_{10}\big[-\ln P_{++}(t)\big]$ in short time for different values of correlation $\tau_c$. Crossover between Markovian and non-Markovian relaxation occurs at $t\sim\tau_c$. The inset shows the decay exponent, $\gamma$.}
\label{fig_3}
\end{center}
\end{figure}

Fig.~\ref{fig_3} depicts the evolution of $P_{++}(t)$ in the log scale for different $g\tau_c$. It shows the same crossover from the non-Markovian to Markovian relaxation when time is comparable to the correlation time. The decay exponent $\gamma$, defined by $P_{++}(t)\approx e^{-(t/T_{1})^{\gamma}}$ can vary from $2$ to $1$ as time increases. Because of this non-Markovian dynamics, the fidelity of the TLS under a coherent control also shares the same feature. We will investigate this more quantitatively in the next subsection.

\subsection{Control fidelity analysis}

The control noise problem is solved by the diagrammatic method in Sec.~\ref{sec_formalism} with TLS initially in the excited state and under a coherent $2\pi$ pulse with truncated Gaussian shape defined by $\Omega(t)=\Omega \ e^{-(t-\frac{t_0}{2})^2/\sigma^2}$ for $0<t<t_{0}$ and zero otherwise. The coherent state amplitudes $\alpha_k$ is related to $\Omega(t)$ by Eq.~(\ref{eq_E(t)}). We take $\sigma=t_0/4$ and use the Gaussian photon DOS given by Eq.~(\ref{eq_rho_omega}).

\begin{figure}[th]
\begin{center}
\includegraphics[angle=0, width=0.9\columnwidth]{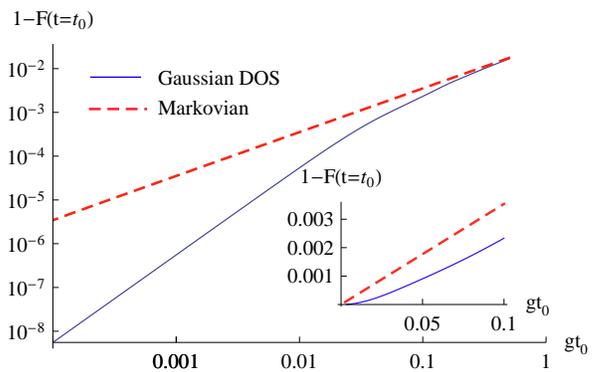}
\caption{(color online). A plot of error, $1-F(t=t_0)$ for a TLS driven by a nominal $2\pi$ Gaussian pulse using a Gaussian DOS with $g\tau_c=0.03$ (solid blue) and a constant DOS (dashed red). For the Gaussian DOS, the $t_0$ dependence of error changes from quadratic to linear when $t_0\sim \tau_c$, whereas the error depends linearly on $t_0$ in the Markovian regime. }
\label{fig_4}
\end{center}
\end{figure}

We evaluate the single contraction diagrams in Fig.~\ref{fig_1}(d) for decoherence of $\sim O[(t/T_1)^\gamma]$. The fidelity is computed by $F(t=t_0)= Tr \left[ P_{ideal} P(t=t_0) \right]$, where $P_{ideal}$ is the ideal reduced density matrix. The error, $1-F(t=t_0)$ is plotted as a function of control duration $g t_0$ in Fig.~\ref{fig_4} and is compared with the Markovian (i.e. broadband) approximation. The third diagram in Fig.~\ref{fig_1}(d) gives a contribution from the contraction between two time lines which tends to be opposite in sign to the dissipation effect of the two graphs with contractions within the the same line. This is evidence of quantum interference between control and dissipation. Owing to the non-Markovian relaxation, the error of the operation goes quadratically with $t_0$ for small $t_0$ and then becomes linear in $t_0$, which is different from the linear $t_0$ dependence in the Markovian limit. The crossover occurs at $t_0\sim \tau_c$ for the same physical reason in the vacuum relaxation process. The fidelity is in the form, $F=1-c(t_0/T_1)^\gamma$, where $1\leq \gamma \leq 2$ and $c$ is a constant.

The relaxation at $t > t_0$ after the pulse finishes is not as simple as the Markovian limit, and is given by
\begin{equation}
P_{\sigma \sigma'}(t>t_0)=P_{\sigma \sigma'}(t=t_0)-f_{\sigma \sigma'}[\mathcal{A}(t)]\times \left[(t-t_0)/T_1 \right]^\gamma.
\end{equation}
The renormalization factor $f_{\sigma \sigma'}[\mathcal{A}(t)]$ is a functional of $\mathcal{A}(t)$, depending on the shape of the pulse. This reflects the history-dependent dynamics of the TLS when we consider a Gaussian photon DOS. Such a feature stands in contrast with the Markovian case, where the function $f_{\sigma \sigma'}[\mathcal{A}(t=t_0)]$ only depends on the total area of the pulse.

\begin{figure}[b]
\begin{center}
\includegraphics[angle=0, width=0.9\columnwidth]{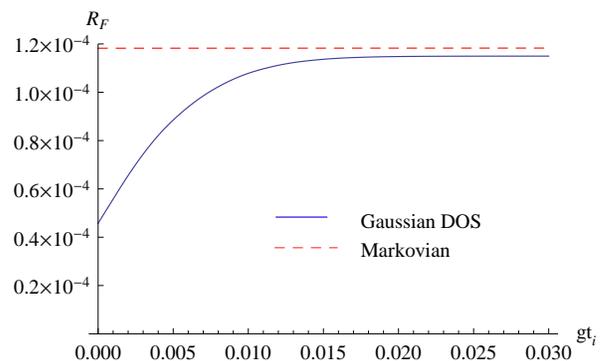}
\caption{(color online). Relative error, $R_F = 1 - F(t_f)/F(t_i)$ as a function of $t_i$ of a TLS under a $2\pi$ Gaussian pulse for fixed $gt_0=g(t_f-t_i)=0.01$ and $g\tau_c=0.01$. For a Gaussian DOS, a change of dependence on $t_i$ occurs when the system evolves from the non-Markovian to Markovian regime.}
\label{fig_5}
\end{center}
\end{figure}

The discussion above is based on a pure starting state. Consider the extension to a mixed starting state. Let the prepared pure state at $t=0$ relax to the mixed state at time $t_i> 0$ when a control operation starts. Then, the TLS is driven by the control pulse and influenced by the vacuum decoherence. The theory is the same except the laser pulse is shifted in time to $\Omega(t) \neq 0 $ in $t_i \leq t \leq t_f$. The fidelity of an example is studied, using a $2\pi$ pulse, which is Gaussian cut off at both ends of the same time interval $\sigma$ from the center as before. The results are given in Fig.~\ref{fig_5} for the relative error defined as the proportionate change of state fidelity between the start and the finish of the pulse, $R_F = 1- [F(t_f) /F(t_i)]$ of the operation as a function of the initial time of the pulse. For a small and fixed duration $t_0=t_f-t_i$, we observe that the relative error increases linearly with $t_i$ and then saturates to a constant after $t_i \sim \tau_c$. This can be roughly understood by employing the decoherence crossover picture discussed in subsection~\ref{subsec_nonMarkov_relaxation}. When the operation is performed inside the non-Markovian region, $F\sim e^{-(gt_i)^2}$ and $R_F \sim 2g^2 t_0 t_i$; whereas in the Markovian limit, $F\sim e^{-t_i/T_1}$ and $R_F \sim t_0/T_1$.

\section{Error checks}\label{sec_comparison}

We  check our method against the exact solution, \cite{shore93} of the driven single-mode JC model for the non-Markovian effects involving control and decay of the TLS and compare with approximations in the master equation approach. It is a limit of the multimode problem for $g\tau_c = \infty$ and $\gamma=2$. We choose four commonly used ME approximations, (i) the Born series, (ii) the Nakajima-Zwanzig (NZ) projection method, (iii) time-convolutionless (TCL) projection\cite{breuer99, breuer02}, and (iv) the additive assumption.\cite{gardiner} Unlike our error bound, these theories contain none. Their results are compared with the exact, the diagrammatic and the classical Rabi solutions. For details of how the first three methods are used in the calculations, see Appendix~\ref{sec_appendix_D}. The additive ME, assumes that the driving terms of the equation of motion by control and by dissipation simply add:
\begin{eqnarray}
\frac{d}{dt}\rho_s(t)=-i[H_{cl}(t),\rho_s(t)]+\hat{K} \rho_s(t),
\label{eq_additive ME}
\end{eqnarray}
where $\rho_s(t)$ is the reduced density matrix of the TLS, $H_{cl}(t)$ the classical control Hamiltonian, and $\hat K$ an appropriate super-operator on $\rho_s(t)$ for the population decay and decoherence effect in the absence of a control (see Appendix~\ref{sec_appendix_D}).

\begin{figure}[tb]
\begin{center}
\includegraphics[angle=0, width=0.9\columnwidth]{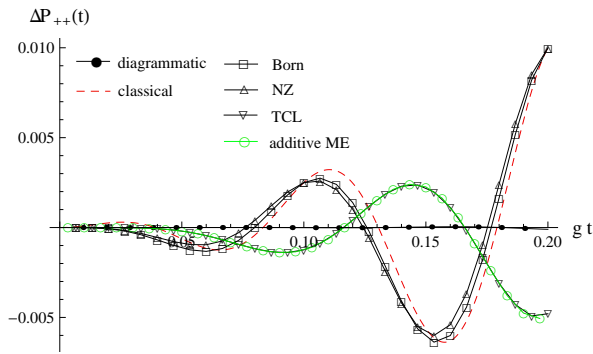}
\caption{(color online) $\Delta P_{++}(t)$, the deviation of $P_{++}(t)$ from the exact solution of a driven single mode JC system using different methods. $\bar n=100 \pi^2$, corresponding to a $4\pi$ rotation at $gt=0.2$. The errors of the ME approximations are as much as that of the classical solution.
}
\label{fig_comparison}
\end{center}
\end{figure}

The TLS is initially in the excited state and driven by a single mode coherent state at resonance with the TLS. The average number of photons, $\bar{n}=100~\pi^2$, is chosen so that, at $gt=0.2$, the area of the classical pulse is $4\pi$. Fig.~\ref{fig_comparison} plots the deviations of the upper state population from the exact solution, $\Delta P_{++}(t) = P_{++}(t)-P^\text{exact}_{++}(t)$, for the six methods above. The diagrammatic method is indistinguishable from the exact solution in the small parameter region of $gt \le 0.2$ with a theoretical error bound $\sim O[(gt)^4]$ and the actual computed results within an error of $10^{-4}$. This error can be further improved to $O[(gt)^{2(n+1)}]$ by including higher order diagrams with $n$ contractions. The Born approximation and the low order NZ give results close to the classical treatment of the electromagnetic fields can be interpreted as a lack of quantum content. The agreement between the low order TCL and the additive approximation might be an indication of the additive nature of TCL. All four ME approximations fail to reproduce the quantum interference effects between control and dissipation which are in the field theoretic treatment. Note that the problems of the NZ and TCL methods are not due to the second order approximation we used because their errors are also of the same order. The failure of the ME methods to account for the high accuracy of control fidelity required by quantum information processing is illustrated by this comparative study.

\section{Conclusion}\label{sec_conclusion}

We have developed a rigorous solution of the multimode Jaynes-Cummings model for a two-level system under the control of a coherent electromagnetic pulse. It treats the control process exactly within a rigorous error bound for the controlled dynamics of the entire system of the small quantum object and the photons to any order in the small parameter in operation time versus the decoherence time, $(t_0/T_2)^\gamma$ where $\gamma$ varies from 1 to 2 as the density of states of the photons varies from flat as in free space to a sharp peak as in high Q cavity. Our theory is quantum in the sense of no stochastic assumption and treats the entire quantum system correctly within any given error bound. A diagrammatic representation provides a simple picture of different physical processes, including the familiar limiting cases of coherent Rabi oscillation and vacuum decoherence. It pinpoints the control-dissipation interference as the quantum effects found by works which precede ours. The effects are relevant to basic quantum phenomena and to technological applications.

This work shows that vacuum relaxation is not the only physical process that results in quantum noise in non-Markovian systems. Fundamental quantum noise due to interference between the control field and relaxation exists and it leads to a decoherence comparable to the vacuum fluctuations. Moreover, our method is not restricted to the Markovian limit and is valid for an arbitrary DOS. Thus, it necessarily goes beyond the optical Bloch analysis.

The time evolution study shows that the relaxation of the TLS can vary from a Gaussian decay to an exponential one, depending on the ratio of time to the correlation time that characterizes the photon DOS. In consequence, the fidelity of a general single qubit operation on a pure state can be cast in the form $F=1-c(t_0/T_2)^\gamma$ for a constant $c$. The Markovian approximation ($\gamma=1$) overestimates the decoherence of the TLS. The Gaussian decoherence ($\gamma=2$) provides a lower bound for the quantum error of a light controlled TLS.

The field theoretic technique provides a completely quantum mechanical description of a small quantum system interacting with the photonic environment. The control noise issue, being important in quantum computing, serves as an concrete example to demonstrate its capability.
This approach may also be applied to other systems under nonclassical photon states. Besides the quantum object, the field theoretic technique also permits a calculation of the physical quantities of the environment, e.g. the quantum feedback on the electric field, photon correlation functions, correlation between the TLS and the photons, etc.

Comparative studies with the existing master equation approximations show their general lack of the quantum effects due to the control-dissipation interplay, which is not restricted to the single-mode model tested. We hope that the results of our rather cursory study of these approximations would stimulate more developments in rigorously bounded approximations for the master equation and further understanding of quantum effects by the contrast and complementarity between the master equation and the field-theoretic approaches.

Our current theory only considers a coherent state with a constant phase. In future work, this can be generalized to describe an ensemble of coherent states with a mixture of phases in order to evaluate the phase error as an extension of previous single mode study.\cite{Gea-Banacloche02}


\begin{acknowledgments}
This research was supported by the U.S. Army Research Office MURI award W911NF0910406. We would like to thank Paul R. Berman and Renbao Liu for helpful discussions. C. K. Chan thanks Wen Yang for a useful comment.
\end{acknowledgments}

\appendix

\section{Rotating wave approximation}\label{sec_appendix_A}

To prove its validity for the control noise problem, we start with the Hamiltonian $H=H_0+V$ in Eq.~(\ref{eq-h}) with the full interaction term, $V=\sum_k g_k (a_k^\dagger+a_k) \sigma_x$. To this, we apply a unitary transformation $e^S$, where, \cite{zheng08}
\begin{eqnarray}
S=\sum_k \frac{g_k}{\omega_k+\omega_0}(a_k^\dagger+a_k) \sigma_x.
\end{eqnarray}
For $g|\alpha|/\omega_0 \ll 1$, we expand the transformed Hamiltonian $\tilde H=e^S H e^{-S}$ up to second order in $g_k$ as:
\begin{eqnarray}
H &=& H_0 + V_\text{RWA}(\{g_k'\}) + V', \\
\text{where} \quad
V' &=& \frac {1}{4\omega_0} \sigma_z \Big[\sum_k g_k'(a_k^\dagger-a_k)\Big]^2.
\end{eqnarray}
$V_\text{RWA}(\{g_k'\})$ is the TLS-field coupling in RWA with $g_k$ replaced by $g_k'= 2 g_k \omega_0/(\omega_0+\omega_k)$. Therefore, the counter-rotating terms leads to an effective perturbation $V'$ as a correction to the RWA. It is an order $O(g\alpha/\omega_0)$ smaller than $V_\text{RWA}$. Hence, in the regime where $1/\omega_0 \ll 1/g|\alpha| \sim t \ll T_1$, the effect of the counter-rotating terms is negligible compared with decoherence.

\section{Application of the Wick's theorem}\label{sec_app_BNEW}

The evaluation of the general perturbation term is simplified by the Wick's theorem, a general bosonic operator defined by $W=\prod_{i}O_{i}$, where $O_{i}=\sum_{k}(u_{ki}b_{k}+v_{ki}b_{k}^{\dagger})$ is a linear combination of bosonic operators $b_{k}$ and $b_{k}^{\dagger}$, and can be rearranged as: \cite{wick50}
\begin{align}
W = &:\hskip -3pt W\hskip -3pt : +\sum_{(i_{1},j_{1})}:W_{i_{1},j_{1}}:\left\langle O_{i_{1}}O_{j_{1}}\right\rangle \nn\\
&+\sum_{(i_{1},j_{1})\neq(i_{2},j_{2})}:W_{i_{1},j_{1},i_{2,}j_{2}}:\left\langle O_{i_{1}}O_{j_{1}}\right\rangle \left\langle O_{i_{2}}O_{j_{2}}\right\rangle\nn \\
&+ \ldots \nn \\
&+\sum_{(i_{1},j_{1})...(i_{n},j_{n})}\left\langle O_{i_{1}}O_{j_{1}}\right\rangle ....\left\langle O_{i_{n}}O_{j_{n}}\right\rangle,
\label{eq_wick}
\end{align}
where $:\hskip -4pt W \hskip -4pt :$ is the normal ordered form of $W$, defined by all the creation operators to the left of the annihilation operators; $:\hskip -4pt W_{j,k,l,\ldots}\hskip -4pt :$ is the normal ordered form of $\prod_{i\neq j,k,l,...}O_{i}$ in which the operators $O_j$ etc. are left out; and the contraction between $O_i$ and $O_j $ is defined only for $i < j$ by a scalar,
\begin{equation}
\langle O_{i}O_{j}\rangle=O_{i}O_{j} - :\hskip -3pt  O_{i}O_{j}\hskip -3pt : ,  \label{eq-contraction}
\end{equation}
Note that the suffix in the $O$ operator denotes its order in the $W$ expression, rather than the time index in the photon operator $A_l$.

\section{Evaluation of control and dissipation}\label{sec_appendix_C}

\begin{figure}[tb]
\begin{center}
\includegraphics[angle=0, width=0.9\columnwidth]{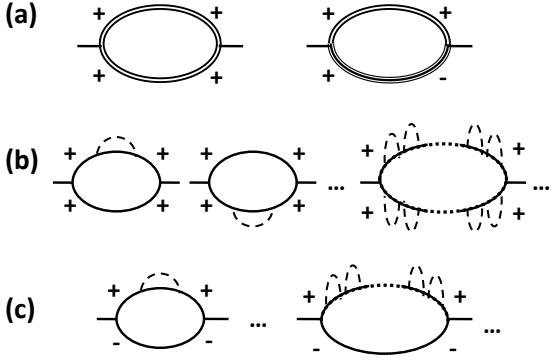}
\caption{(a) gives the diagrams for the classical Rabi solutions for $p_{++,++}(t)$ and $p_{-+,++}(t)$. (b) and (c) provide all the diagrams in the absence of control for $p_{++,++}(t)$ and $p_{-+,-+}(t)$ respectively. In the Markovian limit, the two ends of a dashed line are squeezed to the same point. (b) consists of two lowest order diagrams while there is only one in (c), implying $T_1=T_2/2$ in the Markovian regime.}
\label{fig_8}
\end{center}
\end{figure}

This supplements Sec.~\ref{sec_non-Markovian dynamics} with an analysis of the mathematical structure of the diagrammatic series, two examples of exactly soluble situations, and more details of the evaluation of formulas.

\subsection{The contraction formula}

Central to the vacuum decoherence is the contraction function from Eq.~(\ref{eq-aa}) given by,
\begin{eqnarray}
\langle A(t) A(t')^\dagger \rangle \equiv K(t-t')= \int d\omega \rho(\omega) e^{i(\omega_0-\omega)(t-t')}, \,
\label{eq_contraction}
\end{eqnarray}
where $t$ and $t'$ can be on the same or opposite time lines.

\subsection{Dressed TLS state lines}

A time segment between any two consecutive points chosen from $0, t$ or contracted interaction points in a diagram in Fig.~\ref{fig_1} may be viewed as a dressed TLS line in which all possible interaction terms for the coherent state matrix element are summed. An example term is given by:
\begin{eqnarray}
&&\int^{t_{2n+j+1}}_{t_j} dt_{2n+j}~...~\int^{t_{j+3}}_{t_j} dt_{j+2} \int^{t_{j+2}}_{t_j} dt_{j+1} \nn \\
&&\ \ \times \left\langle \{\alpha\}\right|: A_{2n+j} A_{2n+j-1}^\dagger ... A_{j+2} A_{j+1}^\dagger  :\left|\{\alpha\}\right\rangle
\label{eq_integral_example}
\end{eqnarray}
The normal product in the integrand enables the evaluation of the coherent state matrix element by putting the creation operator to the left with the substitution $a_k^{\dagger}\rightarrow \alpha_k^*$, and the annihilation operator to the right with $a_k\rightarrow \alpha_k$. The control is given by,
\begin{eqnarray}
\sum_{k}\left(g_{k}e^{-i\omega_{k}t}\alpha_{k}+c.c\right) = \frac{\Omega(t)}{2}e^{-i(\omega_{0}t+\phi)}+\text{c.c.},
\label{eq_E(t)}
\end{eqnarray}
which relates $\alpha_k$ with the envelope function $\Omega(t)$, the resonant frequency $\omega_0$ and the phase $\phi$ of the driving field. Thus, an infinite series sum of integrals like Eq.~(\ref{eq_integral_example}) can be carried out, leading to even and odd types of dressed photon lines (represented by Fig.~\ref{fig_1}(c)):
\begin{align}
D_{e} (t , t') &= \cos\left( \frac{\mathcal{A}(t)-\mathcal{A}(t')}{2}\right) \Theta(t-t'),\nn \\
D_{o} (t , t') &= (\pm)i \sin\left( \frac{\mathcal{A}(t)-\mathcal{A}(t')}{2}\right) \Theta(t-t') \; (e^{\pm i\phi}),
\label{eq_dressed_funcyion1}
\end{align}
corresponding, respectively, to the double and triple lines that are dressed by an even and odd number of photons in the control. Note that the dressed function $D_{o}(t,t')$ picks up a $+$ ($-$) sign, when the triple line is on the upper (lower) time line, and gains a phase $e^{i\phi}$ ($e^{-i\phi}$), if the triple line goes from $-$ to $+$ in the clockwise (anticlockwise) sense. Here, $\mathcal{A}(t)=\int_0^t dt'\Omega(t')$ gives the area of the pulse at time $t$. In the absence of control, $D_{e}(t,t')\rightarrow \Theta(t-t')$ and $D_{o}(t,t')\rightarrow 0$, so that the double line is reduced to a single line and diagrams that contain any triple line vanish.

The mathematical expressions representing different diagrams can be obtained by first writing down all the contraction and dressed functions, and then integrating over all time variables corresponding to the vertices of each contraction line. The vertex picks up a factor of $i~(-i)$ if it is on the upper (lower) time line. In the following, we will provide some explicit examples.

The classical Rabi solution corresponds to diagrams with no contraction. The two dressed lines in Fig.~\ref{fig_8}(a) gives the transformation matrices $p_{++;++}(t)$ and $p_{-+;++}(t)$. Using Eqs.~(\ref{eq_dressed_funcyion1}), we have
\begin{eqnarray}
p_{++,++}^{(0)}(t)&=&\cos^{2}\left[\frac{\mathcal{A}(t)}{2}\right],\nn \\
p_{-+,++}^{(0)}(t)&=&-ie^{i(\omega_0 t+\phi)}\cos\left[\frac{\mathcal{A}(t)}{2}\right]\sin\left[\frac{\mathcal{A}(t)}{2}\right].
\end{eqnarray}
which is the Rabi solution without decoherence.

\subsection{Vacuum relaxation of TLS}

In the absence of control, the photon lines are not dressed. A broadband DOS yields the Markovian limit,
\begin{eqnarray}
K^{M}(t-t') = \frac{1}{T_1}\delta(t-t').
\label{eq_contraction_Markov}
\end{eqnarray}
Then all the diagrams in Fig.~\ref{fig_8}(b) and (c) can be summed exactly, yielding respectively:
\begin{eqnarray}
p_{++,++}^{vacuum,M}(t)&=&e^{-t/T_1},\nn \\
p_{-+,-+}^{vacuum,M}(t)&=&e^{i\omega_0 t-t/2T_1}.
\label{eq_spontaneous_Markov}
\end{eqnarray}
This Markovian limit leads to the standard result of spontaneous emission, where $T_2=2T_1$. This relation can also be seen from the lowest order terms in that $p_{++,++}^{vacuum,M}(t)$ contains two lowest order diagrams, while $p_{-+,-+}^{vacuum,M}(t)$ only one.

Using the same argument, for a non-Markovian system with a general DOS and the $\gamma$ parameter defined in Sec.~\ref{subsec_nonMarkov_relaxation}, we have
\begin{eqnarray}
p_{++,++}^{vacuum}(t )&=&1 - \left( \frac{t}{T_1}\right)^\gamma +O\left[\left( \frac{t}{T_1}\right)^{2\gamma} \right],\nn \\
p_{-+,-+}^{vacuum}(t )&=& e^{i\omega_0 t} \left \{1 - \frac{1}{2}\left( \frac{t}{T_1}\right)^\gamma +O\left[\left( \frac{t}{T_1}\right)^{2\gamma} \right] \right\},
\label{eq_spontaneous_Markov2}
\end{eqnarray}
so that $T_2 = 2^{1/\gamma} T_1$.


\subsection{Exact solution to first order in contraction}

Fig.~\ref{fig_1}(d) shows all the diagrams with one contraction with the dressed lines. By the diagrammatic rules, the transformation matrix is:
\begin{align}
& p^{(1)}_{++,++}(t)= 2 Z_1 (t)+Z_2 (t),
\label{eq_control noise} \\
\text{where~}
& Z_1(t) = - D_e(t,0) \int^t_{0} d\tilde t\int^{\tilde{t}}_{0} d\tilde t'
K(\tilde{t}-\tilde{t'}) \nn \\
&\qquad \times D_e(t,\tilde t) D_e(\tilde t,\tilde t') D_e(\tilde t',0), \\
& Z_2(t)=  \int^t_{0} d\tilde t\int^{t}_{0} d\tilde t'
K(\tilde{t}-\tilde{t'}) \nn \\
& \qquad \times  D_o(t,\tilde t) D_o(t,\tilde t') D_e(\tilde t,0) D_e(\tilde t',0). \label{eq_control noise2}
\end{align}
$Z_1$ and $Z_2$ correspond to the non-crossing and crossing diagrams respectively. This result is valid for arbitrary photon DOS and thus covers both the Markovian and non-Markovian regimes. This is the basis for the result in Sec~\ref{sec_non-Markovian dynamics} using a Gaussian DOS, Eq.~(\ref{eq_rho_omega}).

\begin{figure}[tb]
\begin{center}
\includegraphics[angle=0, width=0.9\columnwidth]{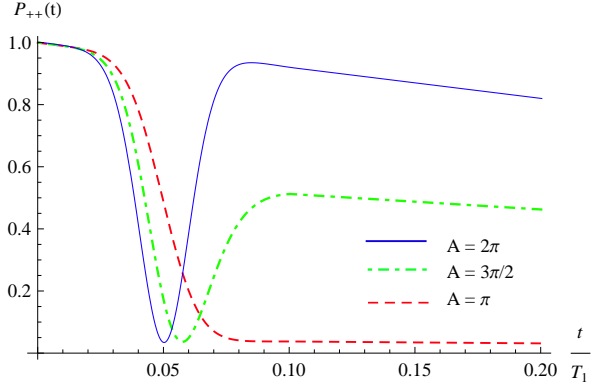}
\caption{(color online). $P_{++}(t)\approx P^{(0)}_{++}(t)+P^{(1)}_{++}(t)$ of an initially excited TLS under a Gaussian pulse in the Markovian limit. It is in quantitative agreement with the optical Bloch solution. }
\label{fig_9}
\end{center}
\end{figure}

The diagrammatic method can reproduce the Markovian results. By Eq.~(\ref{eq_contraction_Markov}) and (\ref{eq_control noise}), Fig.~\ref{fig_9} shows a Markovian example of an initially excited TLS driven by a Gaussian pulse at resonance (the same physical situation as Sec.~\ref{sec_non-Markovian dynamics}~B). Our result not only shows an excellent agreement with the optical Bloch analysis in this Markovian limit, but also allows an understanding of the underlying processes. For instance, the nodes of $P_{++}(t)$ do not reach zero. We note that this feature has been observed in many experiments,\cite{gibbs73,press08,koppens06,stievater01} but cannot exclude other experimental noise sources as the cause of the feature. This theoretical feature can be understood through our diagrammatic representation in Fig.~\ref{fig_1}(d). One can show that the non-zero node arises from the crossing contraction term, which is very different from the vacuum decoherence that contains non-crossing contractions only (Fig.~\ref{fig_8}(b)).


\section{Master Equation Approximations}\label{sec_appendix_D}

The four approximations for the ME approach used in Sec.~\ref{sec_comparison} are detailed here. The standard ME for the TLS up to the second Born approximation is given by (p.~250 of Ref.~\onlinecite{scully97}):
\begin{align}
&\frac{d}{dt}\rho_s^{Born}(t) \notag \\
= &-i \text{Tr}_R\left[ V(t),\rho_s^{Born}(0)\otimes \rho_R  \right]\nn \\
&-\int_0^t dt' \text{Tr}_R \left[ V(t),\left[V(t'),\rho_s^{Born} (t')\otimes \rho_R \right]  \right].
\label{eq_Born}
\end{align}
This equation is derived from a second order expansion of the Liouville equation. In the single mode JC system, at resonance, the interaction is $V(t)=g\left(\sigma_+ a + \sigma_- a^{\dagger} \right)$ and the reservoir density matrix constant, $\rho_R = \left| \alpha  \right\rangle \left\langle \alpha\right|$.

The Nakajima-Zwanzig (NZ) and time convolutionless (TCL) projective operator techniques are outlined systematically.\cite{breuer99,breuer02} The usual assumption of $\text{Tr}_R [ V(t)\otimes \rho_R ]$ vanishing is unnecessary and is not made in our control problem. The second order NZ ME, derived by the method of Breuer \textit{et al.},\cite{breuer99,breuer02} is then:
\begin{align}
&\frac{d}{dt}\rho_s^\text{NZ}(t) \nn \\
= &-i \text{Tr}_R\left[ V(t),\rho_s^\text{NZ}(t)\otimes \rho_R  \right]\nn \\
&+\int_0^t dt' \text{Tr}_R \left[ V(t), \text{Tr}_R \left[V(t'),\rho_s^\text{NZ} (t')\otimes \rho_R \right]\otimes \rho_R  \right]\nn \\
&-\int_0^t dt' \text{Tr}_R \left[ V(t),\left[V(t'),\rho_s^{NZ} (t')\otimes \rho_R \right]  \right].
\label{eq_NZ}
\end{align}
The second order TCL ME shares the same structure but $\rho_s^{NZ} (t')$ is replaced by $\rho_s^{TCL} (t)$ in the integrand.

The additive ME assumes that the control and relaxation terms are additive (see Eq.~(\ref{eq_additive ME})). In the single mode JC system, it becomes:
\begin{align}
 &\frac{d}{dt}\rho^\text{add}_s(t) \label{eq_additive single} \\
= &-i g |\alpha| [\sigma_+ e^{-i\phi}+\sigma_- e^{i\phi},\rho^\text{add}_s(t)]+ g \tan gt  \nn \\
 &\times  \left\{2\sigma_- \rho^\text{add}_s(t) \sigma_+ -\sigma_+\sigma_-\rho^\text{add}_s(t)-\rho^\text{add}_s(t)\sigma_+\sigma_-   \right\}, \nn
\end{align}
where $\alpha = | \alpha| e^{i\phi}$. By neglecting the second term, the first control term leads to the classical Rabi motion. On the other hand, in the absence of the control ($| \alpha|=0$), the second term will produce the vacuum Rabi oscillations of the single mode JC system. In the small time regime ($gt \ll 1$), it corresponds to a Gaussian relaxation.

For an initially excited single mode JC system under a coherent control with $\mathcal{A}=2g| \alpha| t \sim O(1)$, Eqs.~(\ref{eq_Born}-\ref{eq_additive single}) are solved and compared with the exact, the diagrammatic and the classical methods.  The diagrammatic solution is calculated from Appendix~\ref{sec_appendix_C} using the single-mode contraction function $K(t-t')=g^2$. The comparison uses the resultant solutions:
\begin{align}
P_{++}^\text{exact}(t) &= \sum_n  \cos^2\left(g\sqrt{n+1}~t \right)  \frac{| \alpha|^{2n}}{n!}e^{-| \alpha|^2}, \nn \\
P_{++}^\text{diagram}(t) &= \cos^2 g| \alpha| t - \frac{(gt)^2}{4} \cos 2g| \alpha| t   \nn \\
&-\frac{3}{8} \frac{gt}{| \alpha|} \sin 2g| \alpha| t + O\left[ (gt)^4\right],\nn \\
P_{++}^\text{classical}(t) &= \cos^2\left(g| \alpha| t \right).
\end{align}


\end{document}